# Room temperature weak collinear ferrimagnet with symmetry driven, large intrinsic magneto-optic signatures.

F. Johnson[1,a], J. Zázvorka[2,a], L. Beran[2], D. Boldrin[1,3], L. F. Cohen[1,*], J. Zemen[4], M. Veis[2,*]

[1] Department of Physics, Imperial College London, London SW7 2AZ, UK
[2] Charles University, Faculty of Mathematics and Physics, Prague, 121 16, Czech Republic
[3] SUPA, School of Physics and Astronomy, University of Glasgow, Glasgow G12 8QQ, UK
[4] Czech Technical University in Prague, Faculty of Electrical Engineering, Czech Republic
[a] joint first authors
*Corresponding authors

**Abstract**

Here we present a magnetic thin film with a weak ferrimagnetic (FIM) phase above the Néel temperature ($T_N$ = 240 K) and a non-collinear antiferromagnetic (AFM) phase below, exhibiting a small net magnetisation due to strain-associated canting of the magnetic moments. A long-range ordered FIM phase has been predicted in related materials, but without symmetry analysis. We now perform this analysis and use it to calculate the MOKE spectra in AFM and FIM phases. From the good agreement between the form of the measured and predicted MOKE spectra, we propose the AFM and FIM phases share the magnetic space group C2'/m' and that the symmetry driven magneto-optic and magneto-transport properties are maximised at room temperature in the FIM phase due to the non-zero intrinsic Berry phase contribution present in these materials. A room temperature FIM with large optical and transport signatures, as well as sensitivity to lattice strain and magnetic field, has useful prospects for high-speed spintronic applications.

**Introduction**

Antiferromagnets (AFMs) have been of renewed focus for spintronic applications because of the opportunity for high speed, high density memory and logic functions [1,2]. Certain families of AFMs have captured interest because the symmetry of their spin structure offers additional functionality. Semimetallic CuMnAs [3,4] and metallic $Mn_2Au$ [5,6] share the property that each spin sublattice in their collinear spin structure forms inversion partners, which allows for staggered current-induced electrical switching of AFM domains via spin-orbit induced torques. However, although writing information in AFMs has proved encouraging, reading has proved less so, with typically 0.2% anisotropic magneto-resistance (AMR) reported in CuMnAs [7] and 6% in $Mn_2Au$ [6]. Furthermore there have been questions raised regarding influence of anisotropic thermal gradients on the measured signals due to the high current density [please see 8, and the references therein].

AFMs with frustrated exchange interactions between Mn ions also offer unique opportunities for spintronic applications because of the chiral magnetic structures found in both hexagonal crystal symmetry Heuslers $Mn_3X$ [Sn, Ge, Gd] and cubic antiperovskite $Mn_3Y$ [Pt, Ir] and $Mn_3AN$ (Ga, Sn, Ni) compounds [9]. In the latter case, the low temperature non-collinear AFM spin structure in these materials can take two forms, known as $\Gamma^{4g}$ and $\Gamma^{5g}$ related via a 90 degree rotation of the three Mn spins in the (111) plane. For the systems that support the $\Gamma^{4g}$ spin arrangement, the magnetic symmetry along with Weyl points in the band structure close to the Fermi energy allows for non-zero Berry phase curvature underpinning functional properties below the Néel temperature ($T_N$), such as anomalous Hall effect (AHE) [10,11,12,13,14,15,16], anomalous Nernst effect (ANE) [17,18,19], and magneto-optical Kerr effect (MOKE) [20]. The Mn3AN materials have also been shown to support strain sensitive piezomagnetic [21,22,23] or piezospintronic [24, 25, 26] properties.

The attraction of AFMs in terms of their insensitivity to external fields is also a challenge for manipulation of states. Attention has turned recently to exploiting certain classes of ferrimagnets (FIMs) that offer the advantages of speed, and density, with a small unsaturated moment close to their compensation temperature, enabling manipulation of the magnetic properties using small magnetic fields [27]. Previously a collinear FIM phase in Mn3GaN [28] was predicted to occur above $T_N$ in compressively strained films although there no symmetry analysis was made. Following this, a weak magnetism in Mn3NiN grown on SrTiO3 and BaTiO3 substrates above $T_N$ was observed experimentally and attributed to this phase. [29] However the magnetic space group of the latter could not be identified directly. Moreover, Mn3GaN and Mn3NiN are predicted to have different magnetic space groups in the AFM phase, suggesting a direct link between the phase predicted in [28] and the observations in [29] was unlikely.

The symmetry analysis of the FIM phase of compressively strained Mn3NiN, analogous to the FIM phase predicted in Mn3GaN in ref. [28], is conducted in this work using FINDSYM software [34, 35] (see supplementary information for details) and found to be magnetic space group C2'/m'. Using this we determine the spectral dependence of the MOKE for a collinear FIM phase with magnetic space group C2'/m' by spin density functional theory (SDFT) and linear response approximation and compare it to experimentally measured MOKE above and below the Néel temperature on films grown on BaTiO3 substrates. We observe strong qualitative agreement of the data to the theory. We show complementarity in the temperature dependence of MOKE and AHE and that both are maximised at room temperature.

Both intrinsic AHE and MOKE can be observed in materials without net magnetization if a fictitious magnetic field in the momentum space is present [10, 14, 30, 31]. This field is associated with a non-vanishing Berry curvature which is a property of the electronic structure. Within linear response theory, we can express the Anomalous Hall Conductivity (AHC) as an integral of the Berry curvature over the Brillouin zone. [12,14]

$$\sigma_{\alpha,\beta} = \frac{-e^2}{\hbar} \int \frac{dk}{2\pi^3} \sum_{n(occ.)} f[\epsilon(k) - \mu] \Omega_{n,\alpha,\beta}(k),$$

$$\Omega_{n,\alpha,\beta}(k) = -2I \sum_{m \neq n} \frac{\langle km|v_\alpha(k)|kn \rangle \langle kn|v_\beta(k)|km \rangle}{[\epsilon_{kn} - \epsilon_{km}]^2},$$

where $f[\epsilon(k) - \mu]$ denotes the Fermi distribution function with Fermi energy indicated by $\mu$ and $\epsilon_{kn}$ are the energy eigenvalues corresponding to occupied (unoccupied) Bloch band $n$, where $v_\alpha(k)$ corresponds to the velocity operator in Cartesian coordinates. $\Omega_{n,\alpha,\beta}(k)$ is the Berry curvature for given band $n$ in the Brillouin zone.

The Kerr angle ($\theta_K$) and ellipticity ($\eta_K$) relevant for this study (with z-axis perpendicular to film surface) can be calculated from the AHC as follows [32]:

$$\theta_K + i\eta_K = \frac{-\sigma_{xy}}{\sigma_{xx}\sqrt{1 + i(4\pi/\omega)\sigma_{xx}}}$$

Since Berry curvature is a pseudo-vector defined on the Brillouin zone, the presence of AHE and MOKE in systems with time-reversal symmetry broken by the presence of local magnetic moments can be determined by analysing the transformation properties of the Berry curvature pseudo-vector under all the symmetry operations of the particular system.

Unstrained Mn$_3$NiN with the $\Gamma^{4g}$ phase (magnetic symmetry group R-3m' common also to Mn$_3$Ir or Mn$_3$Pt as confirmed by FINDSYM software [34,35]) and with SOC has broken mirror symmetry of the (111) plane $M$, in addition to the broken time reversal symmetry $T$. However, the combined $TM$ symmetry is preserved and makes the Berry curvature an even function of wave vector $k$. [14] After integration over the Brillouin zone, this even property of the Berry curvature makes the AHC as well as MOKE non-zero in the $\Gamma^{4g}$ phase (but zero in the $\Gamma^{5g}$ phase due to the invariance under the mirror symmetry transformations). In-plane biaxial strain *below* T$_N$ induces canting of the Mn moments and lowers the symmetry to magnetic space group C2'/m'. This preserves the non-vanishing Berry curvature and therefore the AHC tensor has two independent elements [33] as shown in Table S1 of the supplementary.

In-plane biaxial strain *above* T$_N$ may induce the collinear FIM phase with net magnetization along the [-1,1,2] axis, analogous to the FIM phase predicted in ref. [28]. Based on symmetry analysis we predict that this novel phase should share the same magnetic space group and form of AHC as the canted $\Gamma^{4g}$ phase [33, 34, 35]. In contrast to a typical collinear FIM system with two mutually compensating sublattices, where MOKE signal vanishes together with net magnetization at compensation temperature (time inversion combined with lattice translation is a symmetry operation so Berry curvature vanishes), the collinear FIM phase (related by a 90° rotation of all moments within the (111) plane to magnetic space group Cmm'm' predicted in Mn$_3$GaN) involves three magnetic sublattices and the magnetic space group C2'/m' enables large Berry curvature despite the nearly compensated net magnetization. We note that biaxial strain applied to a closely related non-collinear system Mn$_3$Pt has been shown to induce a collinear AFM state above room temperature (F-phase) [36]. AHC hosted by the non-collinear phase vanishes in the collinear AFM phase [37] which contrasts with the behaviour of Mn$_3$NiN presented here (the corresponding space groups and AHC tensors are summarised in Table 1 of the Supplementary).

We complement this qualitative analysis of MOKE based on symmetry consideration with simulations of the MOKE spectra using Density Functional Theory (DFT) and linear response theory. Details are given in the Methods sections.

## Experimental and theoretical details

### Measurement of Magnetisation and AHE

100nm Mn$_3$NiN thin films were deposited on (001)-oriented single crystal BaTiO$_3$ (BTO) and SrTiO$_3$ (STO) substrates using pulsed laser deposition from a Mn$_3$NiN stoichiometric target. The substrates were first heated to 400°C and the films were then deposited in a N$_2$ atmosphere, pressure 5 mTorr, using a KrF excimer laser (λ=248nm) with a laser fluency of 0.8 J/cm². The films were then cooled in the N₂ atmosphere to room temperature.

Magnetisation data were collected using the VSM option in a Quantum Design Physical Property Measurement System (PPMS-9T). For measurements with the field out-of-plane, the samples were mounted in a plastic straw with the film surface perpendicular to the applied field. For in-plane measurements, the sample was secured to a quartz paddle with GE 7031 varnish. Linear backgrounds were subtracted. Four-point probe electrical and magneto-transport data were collected in the van der Pauw technique. $\rho_{xy}(H)$ loops were measured in fields of up to 7 T, and the data were then antisymmetrised to extract the Hall component. The saturated values $\rho_{xy,sat}$ were obtained by subtracting the linear ordinary Hall component, which was found by the slope at high field.

## Measurement of MOKE spectra

The MOKE spectra were measured using a field-cooled protocol in a broad spectral region from 1.5 to 5 eV, using a rotating analyser setup [38] with a laser driven Xe lamp as a white light source and a CCD spectrometer as a detector. This allows the experimental data to be collected in the polar configuration (where the magnetic field is perpendicular to the sample surface) with accuracy below 1 mdeg. The sample was measured in an optical cryostat with a static out-of-plane magnetic field of ±200 mT applied using a permanent magnet. The direction of the magnetic field was reversed by mechanical rotation of the permanent magnet. These data were then antisymmetrised to obtain the polar Kerr angle $\theta_K$ = ($\theta_K$(200mT)− $\theta_K$(−200mT))/2. This procedure ensures that the resulting data are of magneto-optical origin, and removes time reversal effects and quadratic magneto-optical contributions. The parasitic Faraday effect from the optical windows of the cryostat was subtracted from the experimental data using the same measurement procedure on a non-magnetic mirror. All MOKE measurements were performed upon cooling and heating with no noticeable temperature hysteresis.

## Theory

We calculate the MOKE spectra using non-collinear spin polarized DFT and linear response theory. In contrast to our earlier work [12] and ref. [31] we have not used the projection on maximally localized Wannier functions here to evaluate the linear response of the systems. Instead, we followed the approach of ref. [30] and used VASP itself to evaluate the elements of the permittivity tensor (which is a counterpart of the optical conductivity tensor):

The Kerr rotation and ellipticity in the polar configuration in the approximation of semi-infinite medium are related to the permittivity as follows:

$$\theta_K + i\eta_K \approx \frac{-\varepsilon_{xy}}{(\varepsilon_0 - 1)\sqrt{\varepsilon_0}}$$

where $\varepsilon_0 = (\varepsilon_{xx} + \varepsilon_{yy})/2$. This approximation is justifiable because of relatively thick samples (100 nm) and their metallic behaviour. This ensures that the light is absorbed within the layer and is no reaching the bottom of the layer.

The projector augmented wave (PAW) method as implemented in the VASP code [39] was used, where the exchange correlation functional is formulated in the generalized gradient approximation (GGA) as parameterized by Perdew-Burke-Ernzerhof [40]. Our results were obtained using a 20 × 20 × 20 k-mesh sampling and a 500-eV energy cut-off to guarantee good convergence. The valence configurations of Mn, Ni, and N, are $3d^64s^1$, $3d^84s^2$, and $2s^22p^3$, respectively.

The intra-atomic Coulomb interaction within GGA was modified through the rotationally invariant approach to GGA+U proposed by Dudarev [41] with U varying from 0 eV to 1.2 eV for Mn site, orbital d. See Supplementary Figure S1 and S2 which show the variation of the MOKE spectra with U and Figure S3 which shows the increase of the size of the local moments with U. In our previous DFT studies we used U = 0 for simplicity. However, the availability of rich experimental data (MOKE spectra) in this study allows us to take the electronic correlations into account and identify the most suitable value of U which we plan to use in our future simulations.

The simulations assume the same c/a ratio as observed in our film experimentally [29]. However, the lattice parameters a and c are rescaled according to the equilibrium lattice parameter obtained by VASP to a = 0.3839 nm and c = 0.3846 nm.

Finally, we note that we use the Gaussian smearing to treat the partial occupancies in k-space integration with a small value of smearing, $\sigma$ = 0.01 eV, which provides us with a detailed view of the Kerr spectra on output.

## Results

An epitaxial (001)-oriented $Mn_3NiN$ thin film, of thickness 100 nm, was grown using pulsed laser deposition at 400 °C on a 10 mm x 10 mm x 0.5 mm single crystal (001)-oriented $BaTiO_3$ substrate, as detailed in the methods section and also described elsewhere [23]. The $c$ and $a$ lattice parameters of the film were measured using neutron diffraction as $c$ = 3.894 Å, $a$ = 3.887 Å, giving a compressive in-plane biaxial strain of $\varepsilon$ = -0.07%. [29] In that work we observed that films grown with compressive strain appeared to support a large AHE above $T_N$. The neutron diffraction data confirmed that the phase above $T_N$ had long-range magnetic order, although the magnetic structure could not be uniquely refined.

Initial characterisation of the magnetic and electrical properties of the 100 nm $Mn_3NiN$ film under investigation is shown in Fig. 1. The Néel temperature $T_N$ = 240 K is determined from the zero-field cooled (ZFC) magnetisation shown in Fig. 1(a). The magnetisation-field loops shown in Fig. 1(c, d, e) at 300 K, 210 K and 170 K respectively show that, on cooling, the film transitions from an isotropic soft FIM phase present above $T_N$ to the AFM $\Gamma^{4g}$ phase with a large coercive field $H_c$. The small finite magnetisation in the AFM $\Gamma^{4g}$ phase originates from strain induced canting of the local moments, creating a net moment in the <112> type directions. In the AFM $\Gamma^{5g}$ phase, this strain induced moment is oriented along <110>, so measurement of the in-plane and out-of-plane moment may give a broad indication of the magnetic state. Indeed, the in-plane and out-of-plane $H_c$ and saturation magnetisation become increasingly different as the film is cooled below $T_N$ - indicative of the transition from FIM to $\Gamma^{4g}$ and finally rotation of the Mn spins moving towards $\Gamma^{5g}$ at the lowest temperatures. We observe that the saturated anomalous Hall resistivity ($\rho_{xy}$) is largest when the film is in the

FIM phase and decreases monotonically as the film is cooled. The origin of this anomalous Hall is from the Berry curvature of the electronic structure, not the small net moment. However, the moment does allow the antiferromagnetic domains to be aligned in an applied magnetic field and therefore facilitates measurement. The magnitude of the longitudinal resistivity ($\rho_{xx}$) for films on BTO compared to films on STO as shown in Fig. 1(f) demonstrates that the films grown on BTO have an additional temperature independent scattering contribution that increases the magnitude of $\rho_{xx}$. This is likely due to the presence of microscopic cracking in the film caused by the phase transitions of the BTO substrate as it is cooled [42].

The polar Kerr rotation spectrum taken at 180 K (when $Mn_3NiN$ is in the $\Gamma^{4g}$ canted AFM phase – see Fig. 2a) is shown in Fig. 2c. The spectral dependence is dominated by a maximum located near 1.7 eV (labelled as *a*) and two minima near 2 and 3.7 eV (labelled as *b* and *c*). At *c* it reaches a maximum absolute value of 12.5 mdeg. Other smaller spectroscopic structures are visible across the spectrum. To aid interpretation of the experimental data, we have calculated the theoretical polar Kerr rotation spectrum using an ab-initio approach (see Methods) including a smearing parameter of 0.01 eV, and the result is displayed as a black line in Fig. 2d. Comparing the initial calculated spectrum with experimental results, the spectral behaviour has some similarities but is shifted towards lower energies.

It has been previously shown that in some collinear antiferromagnets, such as CuMnAs, one can improve the DFT description of the electronic structure by including the on-site Coulomb repulsion using Hubbard parameter U on Mn *3d* orbitals [43]. This repulsion lifts the unoccupied Mn d states further away from the Fermi level, resulting in a blueshift in the optical and magneto-optical response. We have calculated several theoretical spectra of the polar Kerr rotation with U values varying from 0 to 1.2 eV (see Figures S1 and S2 in the Supplementary). We observe a remarkable agreement of the measured and simulated spectra within a narrow range of U values around U = 0.7 eV. The corresponding Kerr rotation spectrum is displayed in Fig. 2d by red line. All spectral structures (*a-c*) in the experimental spectrum were observed in the theoretical spectrum at similar energies. The most prominent feature, the sharp increase in the magnitude of the Kerr rotation at 3.2 eV, coincides with the simulated feature for U = 0.7 eV. It has to be noted that the amplitude of the theoretical spectrum had to be divided by a factor 10 to match the amplitude of the experiment, in agreement with what has been previously found for AHE [12]. This discrepancy can in part be explained by finite temperature and inhomogeneous broadening due to the lack of perfect crystallinity of the films.

We investigated the magneto-optical response in the FIM phase (schematically shown in Fig. 3a) by measuring the polar Kerr rotation at room temperature, and the resulting spectrum is displayed in Fig 3c. The spectrum is significantly different from the spectrum in the non-collinear AFM phase and shows less prominent spectral structures. A local minimum is present near 1.6 eV (labelled as a), otherwise the spectral behaviour is smooth with two minima near 2.8 and 3.5 eV (labelled as b and c). At c, the Kerr angle reaches a maximum absolute value of 18 mdeg.

To understand the origin of the changes in polar Kerr rotation upon the transformation to the FIM phase, we again calculated the theoretical spectral dependence for U=0 eV and for U=0.7 eV from first principles and the results are shown in Fig. 3d. It is still possible to map the observed spectral peaks to the experimental spectrum, but the agreement is notably lower than in the non-collinear AFM phase. We propose two main causes for this: (i) thermal broadening of the transition peaks (the calculations are performed for T = 0 K) and (ii) uncertainty in the direction and magnitude of the magnetic moments, which is challenging to measure in thin films and no similar phase exists in unstrained bulks for comparison. However the energies of all observed structures, except a, are still in reasonable agreement.

## Discussion

In the AFM phase, the magnitude of the Kerr angle (12.5 mdeg) is in the same order as previously reported in bulk $Mn_3Sn$ (20 mdeg) [20] and the spectral dependence of polar Kerr rotation resembles those measured on other compounds containing Mn atoms, such Ni-Mn-Ga [44, 45] or $La_{2/3}Sr_{1/3}MnO_3$ [46, 47, 48]. This suggests that the localized *3d* electrons of Mn are mainly responsible for the magneto-optic (MO) response in the investigated samples. In FIM phase, the maximal absolute amplitude of 18 mdeg at 3.4 eV is even larger than the non-collinear AFM phase (Fig. 3c) and the ratio of Kerr angle to magnetic moment is comparable or larger to other well-known ferrimagnets as shown in Table 1.

To qualitatively link the magneto-optical response of the investigated sample with its electronic structure, the element resolved density of states (DOS) for the *3d* orbitals of Ni and Mn were calculated and are plotted in Fig. 2e and 2f for the AFM phase and Fig. 3e and 3f for the FIM phase. The *s* and *p* DOS are small and too far from the Fermi energy to play an important role in the visible magneto-optical response. Figs. 2f and 3f clearly show a relatively narrow Ni band in the DOS around 1.4 and 1.3 eV below the Fermi energy, while there are almost no states above the Fermi energy. This indicates that the excited states of the electronic transitions, which are responsible for magneto-optical response, may be of Mn origin. Indeed, the DOS of Mn electrons are more complicated with rich energy dependence above Fermi level, as follows from Figs. 2e and 3e. In the AFM phase, there is a gap-like feature above the Fermi energy, from approx. 1.7 to 1.9 eV (marked as yellow in Figs. 2e and 3e). This separates the DOS into two distinct energy regions (marked as I and II in Figs 2b-e). Taking the ground state as *3d* Ni, we can assign all spectral structures seen in the experimental and theoretical spectra to the prominent peaks in the DOS that match in energy, as shown in Fig. 2e. We can therefore infer that the magneto-optical response of $Mn_3NiN$ in the AFM phase is primarily driven by *3d* Ni to Mn charge transfer electron transitions, as schematically shown in Fig. 2b.

Comparing the DOS of the non-collinear AFM phase (shown in Figs. 2e and 2f) to the FIM phase (shown in Figs. 3e and 3f), there is very little difference between the Ni states but large differences between the Mn DOS. This is as expected since only the Mn magnetic moment alignment is different in both phases. There is clearly a rearrangement of valence electronic states from $\Gamma^{4g}$ to FIM. The gap-like feature, highlighted in yellow, has completely disappeared in the FIM phase and instead there is a large band in Mn DOS 1.9 eV above the Fermi energy. This is a clear origin of some of the differences between the polar Kerr rotation spectra of the FIM and AFM phases. Focusing now on the FIM phase, we identified two prominent bands in Mn DOS above the Fermi energy as correlating to the spectral features a and b present in both the experimental data and theoretical calculations (Figs. 3c and 3d). These bands are likely to be considerably broadened at room temperature giving a smoother magneto-optical response than zero-temperature calculations predict. The schematic diagram of the transition mechanism in FIM phase is shown in Fig. 3b.

Since the Mn$_3$NiN non-collinear AFM and FIM phases can be clearly identified from the MO response, we have measured the temperature dependence of the polar Kerr rotation spectra across the transition from 180K to 270K. The temperature evolution of the spectra is shown in Fig. 4b. From the figure one can see that notable changes in the spectra with temperature occur within the spectral range from 2 to 4 eV, reaching their maxima in the vicinity of 3eV, which is consistent with our previous discussion. This is made even clearer by observing the difference spectrum between 180K and 300K in Fig. 4c, with the gap region highlighted in yellow.

The change in the MO response occurs within relatively broad temperature range, indicating a gradual transformation of the samples surface from non-collinear AFM to FIM phase upon heating. As the incident light has a spot size 2mm in diameter, the MOKE spectrum carries integral information over a large region of the film. In collinear antiferromagnets, previous reports have indicated that the transition across the Néel temperature results in a non-homogeneous magnetic order in the sample, before reaching the single magnetic phase [49]. We therefore took an intermediate polar Kerr rotation spectrum at 210K and calculated the theoretical spectrum assuming a phase mixture of 40% FIM and 60% non-collinear AFM. The result is shown in Fig. 4d where the calculated mixed spectrum has good agreement with the experimental data. Further investigation of the transition in the future would be facilitated using a scanning MOKE microscope with a focused beam of light, energy 3eV, which could laterally resolve the magnetic order in the sample.

It is well understood that the intrinsic AHC $\sigma_{xy}$ is proportional to the integrated Berry phase curvature across the Brillouin-zone, while the magnitude of the MOKE signal relates to both the magnitude of the Berry phase curvature and the specific allowed optical transitions. In Mn$_3$NiN, the maximum MOKE and AHE signals are both found in the FIM phase that survives robustly at room temperature. The Berry curvature is also predicted to be a factor of two higher in the FIM phase due to the changes in band structure. Fig. 4e shows the direct comparison between the temperature dependence of the integrated MOKE spectra (in the spectral range from 1.5 to 4.3 eV) and $\sigma_{xy}$, with striking similarity. Clearly the temperature dependence of the Berry phase - which is predicted and found to be maximal in the FIM phase and zero in the $\Gamma^{5g}$ AFM phase - dominates the temperature dependence of both phenomena.

## Conclusions

In this study we use MOKE theoretical analysis and experimentation to confirm the magnetic space group of the high temperature FIM phase of Mn$_3$NiN thin film on BTO. We show that the intrinsic properties of the band structure (non-zero Berry curvature) dominates the large MOKE response in the soft ferrimagnetic phase at room temperature. The material family overall demonstrates large magneto-optical signals and anomalous Hall effect, and very early steps in piezoelectric control of the AFM phase have recently been demonstrated [50]. A magnetic symmetry driven FIM phase which may support fast magnetisation dynamics, has wide ranging implications for spintronic application as well as for example high frequency magneto-optical spatial light modulations [51].The work is also timely in light of the recent classification of magnetic materials, named 'altermagnets' that are also collinear spin systems, and like the ferrimagnetic phase studied here, show the AHE without strong relativistic effects [52, 53].

# Figures

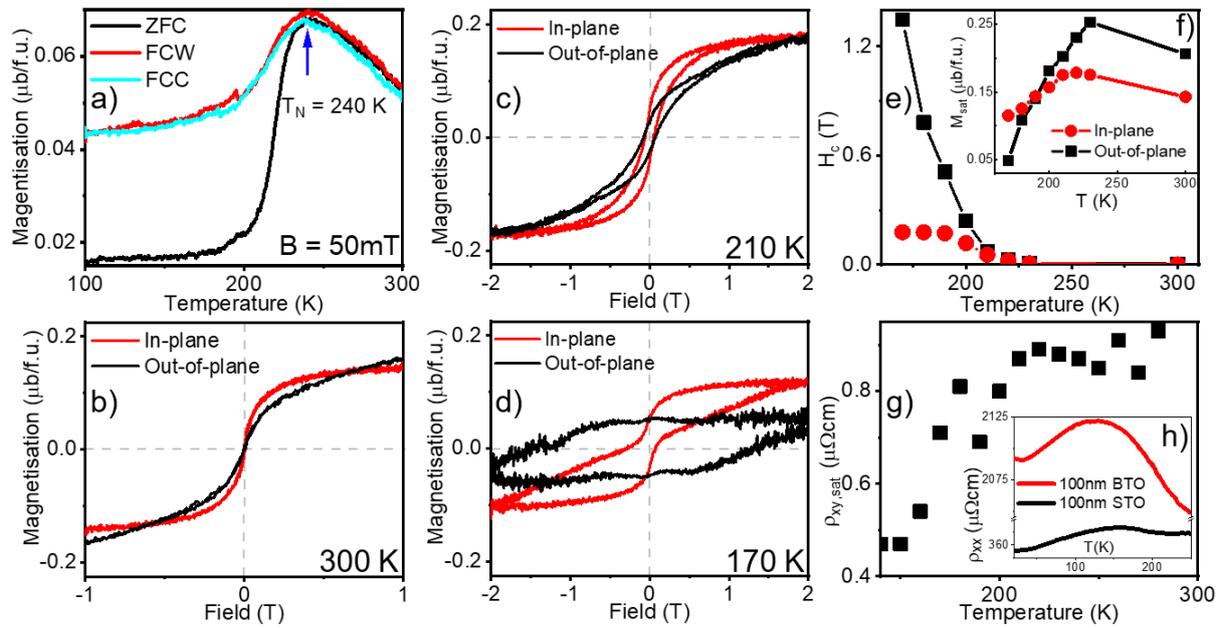

Figure 1: (a) The antiferromagnetic transition temperature $T_N$ = 240K identified from field-cooled (measured on both warming and cooling) and zero-field cooled M(T) in 50mT. M(H) is shown in (b)(c) and (d) at 300K 210K and 170K respectively. (e) The coercive field $H_c$(T) for field in the plane of the film and out-of-the plane and the inset. (f) showing the saturation magnetisation in the two field directions. (g) The saturated anomalous Hall resistivity $\rho_{xy,sat}$(T) and the inset (h) shows the longitudinal resistivity $\rho_{xx}$(T) of the Mn$_3$NiN film on BTO and STO substrates. $\rho_{xy,sat}$(T) is obtained by subtracting the ordinary Hall contribution.

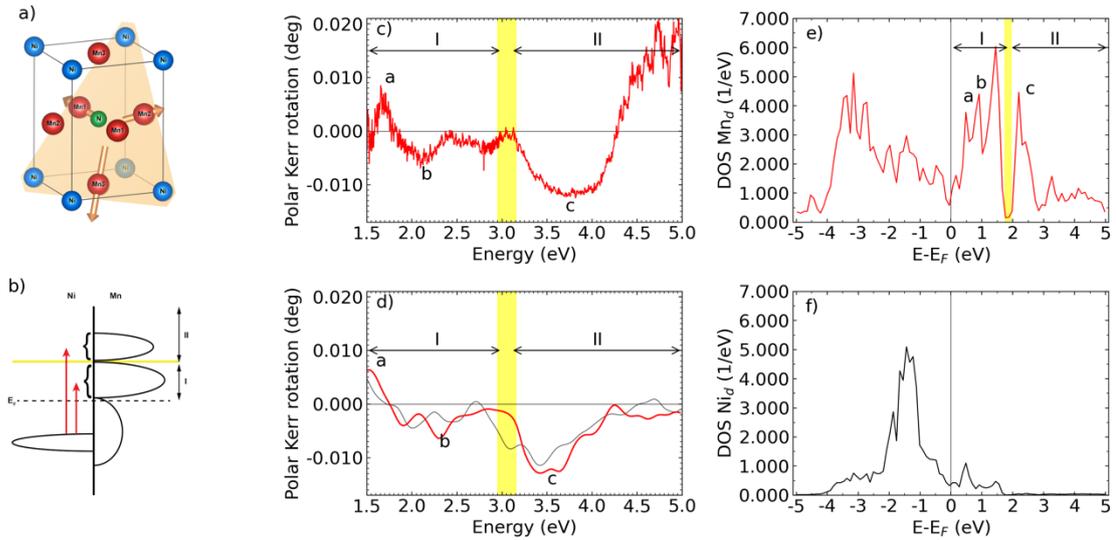

Fig 2. (a) Schematic picture of the unit cell of Mn$_3$NiN together with the Mn magnetic moment alignment in non-collinear AFM phase (b) Simplified picture of the magneto-optically active charge transfer electronic transitions from Ni to Mn *3d* states. The arrows with the curly backets represent groups of transitions for I and II energy regions observed in Mn DOS (c) Experimental spectra of polar MOKE rotation taken at 180 K, obtained from field-cooled measurements in +/- 200 mT. The yellow region is the position of the gap-like feature in Mn DOS taken from Ni states (d) Ab-initio calculated polar MOKE spectrum for AFM phase and U=0 (black) and 0.7 eV (red) – theory multiplied by 0.1. (e) Element resolved DOS for Mn *3d* states with marked gap-like feature (f) Element resolved DOS for Ni *3d* states.

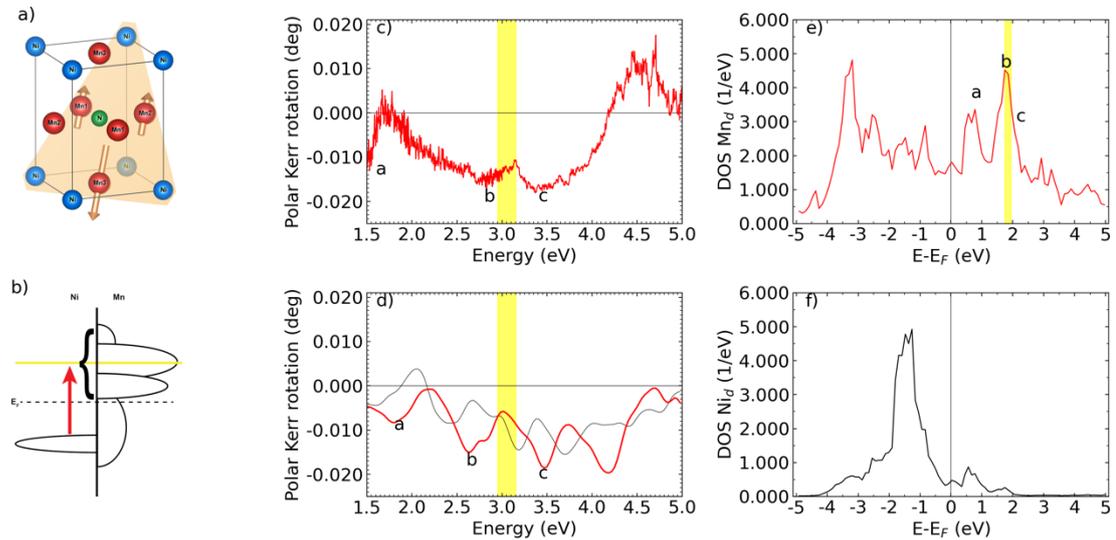

Fig 3. (a) Schematic picture of the unit cell of Mn3NiN together with the Mn magnetic moment alignment in FIM phase (b) Simplified picture of the magneto-optically active charge transfer electronic transitions from Ni to Mn *3d* states. The arrow with the curly backet represents all transitions for the whole investigated energy region (c) Experimental spectra of polar MOKE rotation taken at 300 K, obtained from data taken at +/- 200 mT. The yellow region is the position of the gap-like feature in Mn DOS of AFM phase taken from Ni states (d) Ab-initio calculated polar MOKE spectrum for FIM phase and U=0 (black) and 0.7

eV (red), theory multiplied by 0.1. (e) Element resolved DOS for Mn *3d* states with marked gap-like feature in AFM phase (f) Element resolved DOS for Ni *3d* states

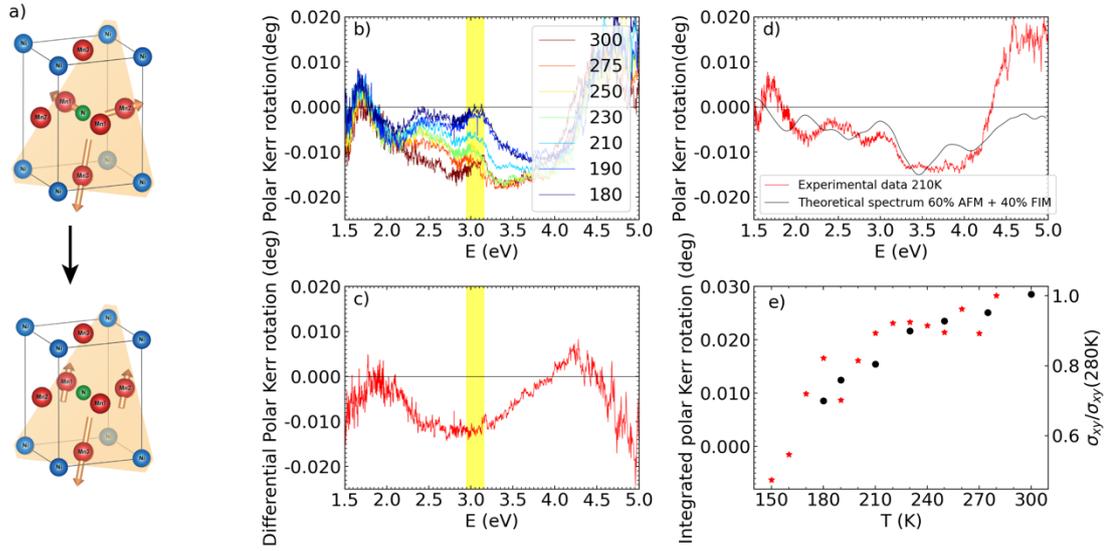

Fig 4. (a) Schematic picture of the unit cell of Mn3NiN together with the Mn magnetic moment alignment across the transition from AFM to FIM phase induced by the change of temperature (b) Temperature evolution of experimental spectra of polar MOKE rotation, obtained from data taken at +/- 200 mT. The yellow region is the position of the gap-like feature in Mn DOS of AFM phase taken from Ni states (c) Differential MOKE spectrum at 300K-180K with marked position of gap-like feature in Mn DOS of AFM phase (d) Experimental MOKE spectrum taken at 210K together with theoretical calculation when taking into account 60% of AFM and 40% of FIM phase – theory is multiplied by 0.1 (e) A comparison of the temperature evolution of AHE conductivity (red stars) and integrated MOKE (black circles) between 1.5 and 4.3 eV.

**Table 1:** A comparison of the anomalous Hall effects present in a variety of magnetic materials. The strength of the AHE relative to the magnetization of the material is parametrised by the ratio column.

| Material | Magnetic ordering | MOKE Amplitude (mDeg) | Magnetisation ($\mu_B$/f.u.) | Ratio | Origin | Reference |
|---|---|---|---|---|---|---|
| $Mn_3NiN$ (100 nm on BTO) at 300K | FIM | 18 | 0,08 | 225 | This work | This work |
| $Y_3Fe_5O_{12}$ | FM | 150 | 5 | 30 | Theory | [54] |
| $Bi_3Fe_5O_{12}$ | FM | 1100 | 4,4 | 250 | Theory | [54] |
| CoPt | FM | 500-1000 | 2,2 | 227 | Theory + exp. | [55] |
| $Fe_3GeTe_2$ | FM | 1000-3000 | 6,3 | 159 | Theory | [56] |
| AuMnSb | FM | 700 | 4,2 | 167 | Theory + exp. | [57] |
| PtMnSb | FM | 1800 | 3,97 | 453 | Theory + exp. | [58] |

| Mn$_3$Sn (bulk) | AFM | 20 | 0,005 | 4000 | Experiment | [20] |
| NiMnSb | FM | 1200 | 3,85 | 312 | Experiment | [58] |

## Acknowledgements

We would like to acknowledge fruitful discussions with Jakub Železný and Joerg Wunderlich.
The work was partially supported by Ministry of Education, Youth and Sports of the Czech Republic from the OP RDE programme under the project MATFUN CZ.02.1.01/0.0/0.0/15_003/0000487, and Czech grant agency (grants No. 19-09882S and No. 22-21974S). J. Zazvorka acknowledges support from Charles University (Grant No. PRIMUS/20/SCI/018). The work of JZ was supported by the Ministry of Education, Youth and Sports of the Czech Republic from the OP RDE programme under the project International Mobility of Researchers MSCAIF at CTU No. CZ.02.2.69/0.0/0.0/18 070/0010457, and through the e-INFRA CZ (ID:90140).
F.J acknowledges funding from Hitachi Cambridge and F.J and LFC from the UK Engineering and Physical Sciences Research Council (EPSRC).
D.B. is grateful for support from a Leverhulme Trust Early Career Fellowship (No. ECF-2019-351) and a University of Glasgow Lord Kelvin Adam Smith Fellowship.